\documentstyle[sprocl,epsf]{article}

\def\Journal#1#2#3#4{{#1} {\bf #2}, #3 (#4)}

\def\NPB{{\em Nucl. Phys.} B}
\def\PLB{{\em Phys. Lett.}  B}

\def\ZPC{{\em Z. Phys.} C}
\def\PRep{{\em Phys. Rep.}}
\def\JETP{{\em Sov. Phys. JETP}}
\def\SJNP{{\em Sov. Jour. Nucl. Phys}}
\def\IJMP{{\em Int. J. Mod. Phys.} A}

\def\be{\begin{equation}}
\def\ee{\end{equation}}
\def\bea{\begin{eqnarray}}
\def\eea{\end{eqnarray}}
\def\Pom{{\bf I\!P}}
\def\xpom{{x_\Pom}}
\begin{document}

\title{SECONDARY REGGEONS IN DIFFRACTIVE DEEP INELASTIC SCATTERING
- THE MICROSCOPIC QCD EVALUATION}

\author{W. SCH\"AFER}

\address{Institut f\"ur Kernphysik, Forschungszentrum J\"ulich,
D-52425 J\"ulich\\E-mail: Wo.Schaefer@fz-juelich.de}


\maketitle\abstracts{ We present the microscopic QCD
evaluation of the secondary reggeon contribution to
diffractive DIS. It is shown that the interference between
Pomeron and Reggeon enters in the maximal possible way.
In the region of large $\beta \simeq 1$ it is shown, that
the structure functions of the Reggeon and Pomeron as well as
the Reggeon-Pomeron interference SF have a universal
$\sim (1-\beta)^2$ behaviour.}

The secondary Regge trajectories $(f,\omega,\rho,A_2)$ are well known from
the phenomenology of soft hadronic reactions at high energies.
For example the cross sections for processes that involve the 
exchange of non-vacuum quantum numbers are governed by the 
t-channel reggeon-exchanges and  exhibit an energy dependence like
$\left. d \sigma/ dt \right |_{t=0}  \propto (s /s_0)^{2\Delta_R} $,
with an almost universal intercept $\Delta_R \simeq -0.5$.
\begin{figure}[h]
%\begin{center}
\epsfxsize=1.0\hsize
\epsfbox{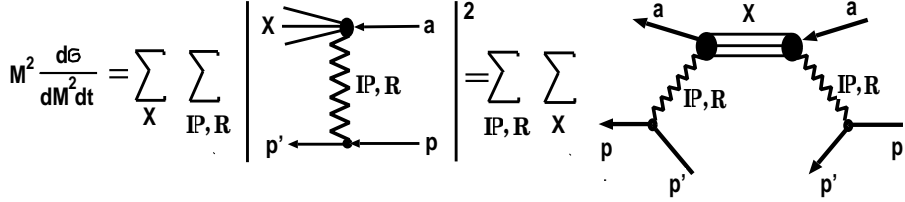}
\caption{Regge expansion of the inclusive cross section for $ap \to p' X$, M is the invariant
mass of the inelastically excited state $X$.}
 \label{fig1}
%\end{center}
\end{figure}
 Hadronic
total cross sections are well described by Regge-parametrizations of the
form $ \sigma_{tot} (ab) = \sigma_{\Pom} (s/s_0)^{\epsilon_{\Pom}} 
+  \sigma_{R} (s/s_0)^{\epsilon_{R}}$
with $\epsilon_R \simeq -0.5, \epsilon_\Pom \simeq 0.1$. Also for the
inclusive hadronic reactions there exists a successful Regge-phenomenology
\cite{Triple,Kaidalov}, which has recently been extended
to diffractive DIS\cite{Triple-prep,H1}. For diffraction dissociation of the
hadronic projectile $ap \to pX$ (see fig 1a) the large Regge parameter 
is $1/x_\Pom = s/M^2$. As can be seen from the diagram in fig.1 there 
appear the forward scattering amplitudes for $a\Pom \to a\Pom, aR\to aR$,
as well as an interference amplitude $a\Pom \to aR$. Schematically one
can write (we do not show explicitly the reggeon-residue functions and 
signature factors):
\bea
M^2 {d\sigma \over dM^2 dt} = {\sigma^{pp}_{tot}\over 16 \pi}
\cdot \left[ \sigma_{tot}(a\Pom)\cdot({1\over x_\Pom})^{2\epsilon_\Pom}
+ 2\Sigma(a\Pom \to a R)\cdot ({1\over x_\Pom})^{\epsilon_\Pom+\Delta_R}\right.
\nonumber \\ +\left. \sigma_{tot}(aR)\cdot({1\over x_\Pom})^{2\Delta_R}\right] 
\propto ({1\over x_\Pom})^{2\Delta_{eff}} \, . \nonumber
\eea
In DIS the projectile $a$ is the virtual photon $\gamma^*$ and 
one readily introduces the pomeron and reggeon structure functions
$ F_{2\Pom,R}(x,Q^2) = Q^2/(4\pi^2 \alpha_{em}) \cdot \sigma (\gamma^* \Pom,R)$ 
and the interference structure function 
$ F_{2R\Pom}(x,Q^2) = Q^2/(4\pi^2 \alpha_{em}) \cdot \Sigma (\gamma^* R \to \gamma^* \Pom)$.
%In the expansion given above the $\xpom$-dependence 
%is determined by the relevant intercepts: for asymptotically  
%small $\xpom$ the vacuum exchange dominates and yields an about flat
%cross section, 
For $\xpom > 0.1$ the secondary trajectories take over, and 
the effective intercept $\Delta_{eff}$, that parametrizes the local 
$\xpom$ behaviour approaches the pomeron
intercept for $\xpom \to 0$ and decreases in the region of large $\xpom > 0.1$ 
where the secondary exchanges dominate. Such a behaviour has been seen by H1
in diffractive DIS \cite{H1expo,Kolya}. 
Not much is known in the literature 
about the importance of the pomeron-reggeon interference term: Triple-Regge
fits to hadronic data are not conclusive \cite{Triple}. If one
adopts the naive point of view that one can attribute the hadronic
(particle-) state vectors to the pomeron and reggeon one
may be mislead to conclude that such interference amplitudes 
should vanish. For the orthogonality of state vectors  $\langle R|\Pom\rangle\ = 0$
implies vanishing parton number and momentum integrals \cite{Triple-prep,Interference}, 
which suggests a strong suppression
of such interference SF's. However we shall demonstrate, that such 
a reasoning is not born out by pQCD.
%\section{Reggeons in DIS}

The familiar decomposition of structure functions into sea and valence 
quark contributions $F_2 (x,Q^2) = F_{sea}(x,Q^2) + F_{val}(x,Q^2)$
serves to identify the Reggeon and Pomeron contributions 
%(see fig \ref{fig2})
to the total $\gamma^*$--absorption cross section (the inclusive SF).
At moderately small $x$ the standard fits give
$F_{val}(x,Q^2) \propto (1/x)^{\gamma}$ with $\gamma \sim -0.45$ 
whereas the sea SF shows a low-$x$ behaviour
$F_{val}(x,Q^2) \propto (1/x)^{\Delta_\Pom(Q^2)}$ with $\Delta_\Pom(Q^2) 
\sim 0.1 - 0.4$, the intercepts being thus compatible with the hadronic 
phenomenology. To each pQCD diagram that contributes to the total cross section 
there is one for the diffractive amplitude as indicated in fig.2.; 
the building block
for the pomeron contribution is the gluon structure function of
the target $d G(\xpom, Q^2)/d\ln Q^2$ \cite{Pomeron}. The analogous
quantity for the reggeon contribution is the $\bar{q} q$ annihilation
amplitude $A(s',k^2)$ as shown in fig.2a). This amplitude has been studied 
extensively in the literature \cite{Amplitude}. For the illustrative purposes
it is sufficient to know that in the DGLAP approximation $A(x,Q^2) \simeq
8 \pi^2 dv(x,Q^2)/d\ln Q^2$, where $v(x,Q^2)$ is the target's valence quark
distribution. 
% The important point is that the gluon and valence structure function
%enter the calculation at the hardness scale $\bar{Q}^2 = (m_q^2 
%+ \vec{k}^2_\perp)/(1-\beta)$ We shall further concentrate on the large-$\beta$ 
%limit where we are safe in the pQCD domain. 
%\begin{figure}[h]
%\begin{center}
%\epsfxsize=1.0\hsize
%\epsfbox{sigmatotal-new.eps}
%\caption{a): DIS on valence quarks can be identified with the t-channel reggeon
%exchange contribution to the SF. b): DIS on sea quarks corresponds to the Pomeron
%exchange. Dashed lines are gluons, solid lines represent quarks.}
% \label{fig2}
%\end{center}
%\end{figure}
\begin{figure}[h]
%\begin{center}
\epsfxsize=1.0\hsize
\epsfbox{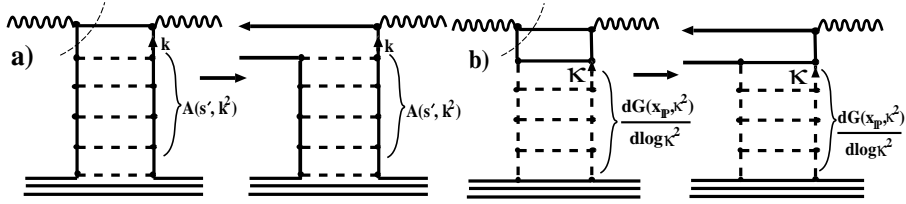}
\caption{The contributions for diffractive excitation of the $q \bar{q}$ state which
dominate the large--$\beta$ region. a) inclusive DIS on a valence quark and the 
corresponding diagram for the diffractive amplitude (reggeon exchange); b) the same for
the sea quark contribution(pomeron exchange)}
 \label{fig3}
\end{figure}
\begin{figure}[h]
%\begin{center}
\epsfxsize=1.0\hsize
\epsfbox{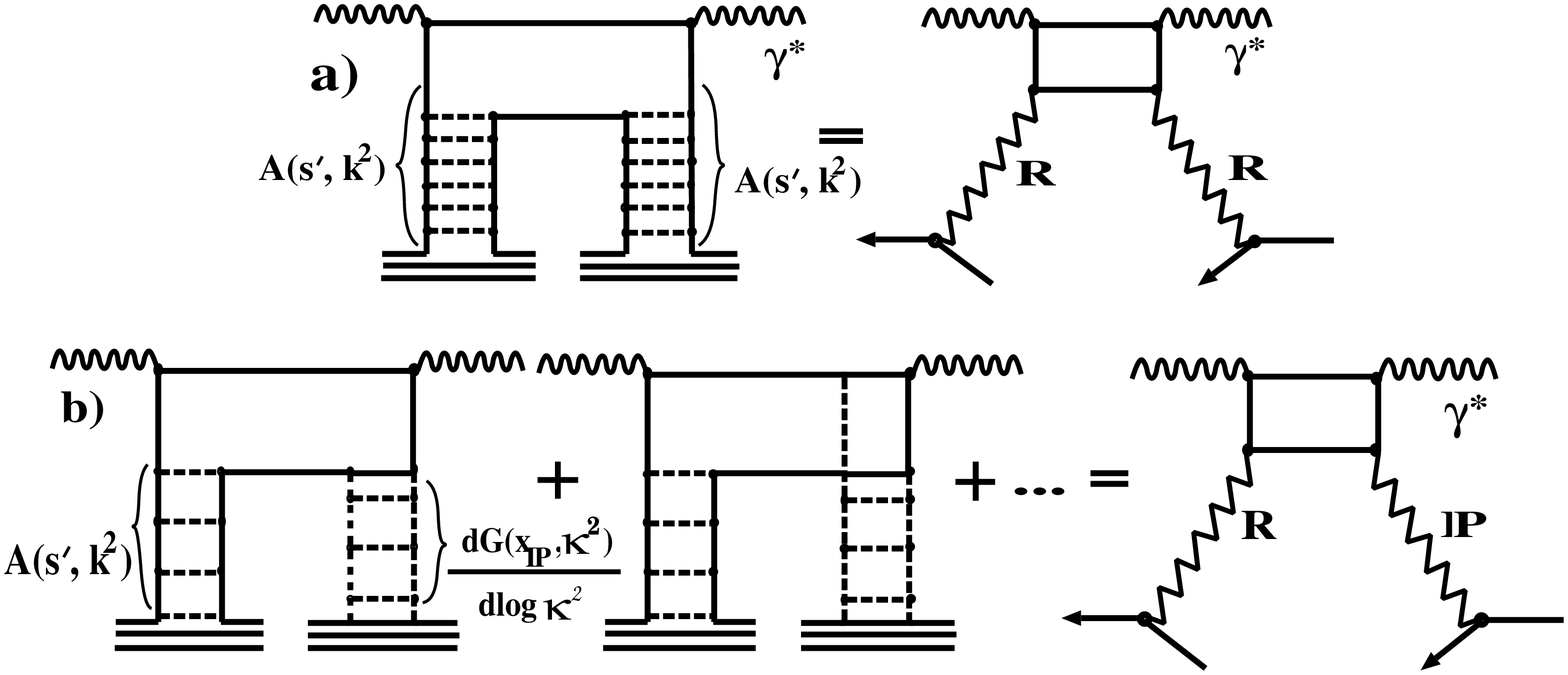}
\caption{Diagrams for the diffractive cross section: a) RR-contribution, b)R$\Pom$-interference.
Not shown is the $\Pom \Pom$ term that arises from squaring the amplitude in fig2 b).}
\label{fig4}
\end{figure}
One may even go further and use an approximate DGLAP formula for the low $\xpom$--limit
$d v(\xpom, \bar{Q}^2)/d \ln Q^2 = C_F \alpha_S(Q^2)/(2\pi\alpha_R)  v(\xpom, Q^2)$
The final result \cite{valence} for the diffractive structure function shall then take the symmetric form:
\bea
F_2^{D(3)}(\xpom,\beta,Q^2) \propto \beta (1-\beta)^2 \alpha^2_S(\bar{Q}^2)
\left| \eta_\Pom G(\xpom,\bar{Q}^2) + \eta_R {C_F \over 2\pi \alpha_R}\xpom v(\xpom,\bar{Q}^2) \right|^2 \nonumber,
\eea
where the hardness scale  $\overline{Q}^2 = (m_q^2 + \vec{k}^2_\perp)/(1-\beta)$
and for large $\beta$ one is safe in the pQCD domain \cite{HT}.
Let us discuss the salient features of this equation: first we see now
explicitly that there are no extra suppression factors whatsoever for the 
$R \Pom$--interference. It enters in the \emph{maximal possible way}.
Second, the $\beta$--dependence is a universal $(1-\beta)^2$ for the pomeron, 
reggeon and the interference structure function. 
In particular, the reggeon structure function differs from the common theoretical
guesses $F_{2\pi} \sim (1-\beta)$ for the pion structure function in the large $\beta$ region.
Third we encountered the manifest leading twist effect. For the longitudinal photons
there will emerge the twist four effect in much the same way as it was
discussed first by Genovese et al.\cite{HT} for the pomeron. Fourth our knowledge of
the valence and gluon structure functions implies that
the result is in the correct order in magnitude with what can be expected from the
analysis of the H1 data \cite{Triple-prep,H1}.

A final comment on possible applications is in order: the theoretical
understanding of the reggeon and inteference contributions has important
impact on the theory of the nuclear shadowing of the valence quark densities,
a topic which has not been adressed before. Vice versa the  
phenomenology of nuclear shadowing puts much constraint on the diffractive 
production mechanisms. For instance we have found that the much discussed
hard gluon dominated pomeron favoured by the DGLAP fits to the
H1 data yields an unacceptably large shadowing
of the deuteron gluon density \cite{deuteron}. A different option
for the pomeron structure function where quarks and gluons share the momentum of
the pomeron in about equal fractions, as it emerges from the colour dipole calculations of Nikolaev and Zakharov, (which are incorporated in the above 
presented evaluations of the pomeron contribution)  
gives the correct amount of gluon shadowing.

\end{document}